\documentclass[fleqn,usenatbib]{rasti}

\usepackage{newtxtext,newtxmath}

\usepackage[T1]{fontenc}

\DeclareRobustCommand{\VAN}[3]{#2}
\let\VANthebibliography\thebibliography
\def\thebibliography{\DeclareRobustCommand{\VAN}[3]{##3}\VANthebibliography}

\usepackage{graphicx}	
\usepackage{amsmath}	
\usepackage{bm}
\usepackage{enumitem}
\usepackage{orcidlink}
\usepackage{pifont}
\newcommand{\cmark}{\ding{51}}%
\newcommand{\xmark}{\ding{55}}%
\newcommand{\ignoreme}[1]{}
\usepackage{hyperref}

\newcommand{\md}{{\mathrm{d}}}

\title[\texttt{cuDART}: GPU Raytracing in Special Relativity]{\texttt{cuDART}: a GPU-accelerated ray tracing code for generating synthetic observations of relativistic astrophysical sources}

\author[H. Whitehead et al.]{
Henry Whitehead$^{\orcidlink{0009-0006-0716-0965}}$$^{1,2}$\thanks{E-mail: henry.whitehead@ist.ac.at}, Emma L. Elley$^{\orcidlink{0009-0002-5349-908X}}$$^{2}$, Christopher N. Everett$^{\orcidlink{0000-0001-5181-108X}}$$^{2}$.
\\
$^{1}$Institute of Science and Technology Austria (ISTA), Am Campus 1, Klosterneuburg A-3400, Austria \\
$^{2}$Department of Physics, Astrophysics, University of Oxford, Denys Wilkinson Building, Keble Road, Oxford OX1 3RH, UK\\
}

\date{Accepted XXX. Received YYY; in original form ZZZ}

\pubyear{\the\year{}}

\begin{document}
\label{firstpage}
\pagerange{\pageref{firstpage}--\pageref{lastpage}}
\maketitle

\begin{abstract}
The generation of synthetic observations is required to compare simulations of astrophysical phenomena with real sources. We present a new open-source toolkit \texttt{cuDART} which allows meshed simulation data to be rapidly rendered into synthetic observations of optically thin emission, automatically accounting for a range of relativistic and geometric effects such as Doppler boosting. Unlike previous visualisation schemes which generally consider a single snapshot in time, \texttt{cuDART} is capable of accounting for the finite time delay across astrophysical distances by rendering data from multiple simulation epochs simultaneously. We demonstrate how failing to account for a finite light time delay can result in synthetic observations with incorrect observed motion, morphologies and fluxes. The code does not require a static pattern frame for integration, relaxing standard assumptions required for analytical observational estimates. The rendering processes is accelerated through hardware (using GPUs) and software (using the 3D differential digital analyser algorithm), allowing for rapid generation of  observations from a large number of viewing orientations. In this paper we describe the physical motivation, document the code operation and test the output against theoretical relativistic and geometric expectations. \texttt{cuDART} is publicly available via \href{https://github.com/hwhitehead/cuDART}{\texttt{GitHub}} and documented using \href{https://cudart.readthedocs.io/en/latest/index.html}{ReadTheDocs}.
\end{abstract}




\section{Introduction}

\begin{figure*}
    \centering
    \includegraphics[width=2\columnwidth]{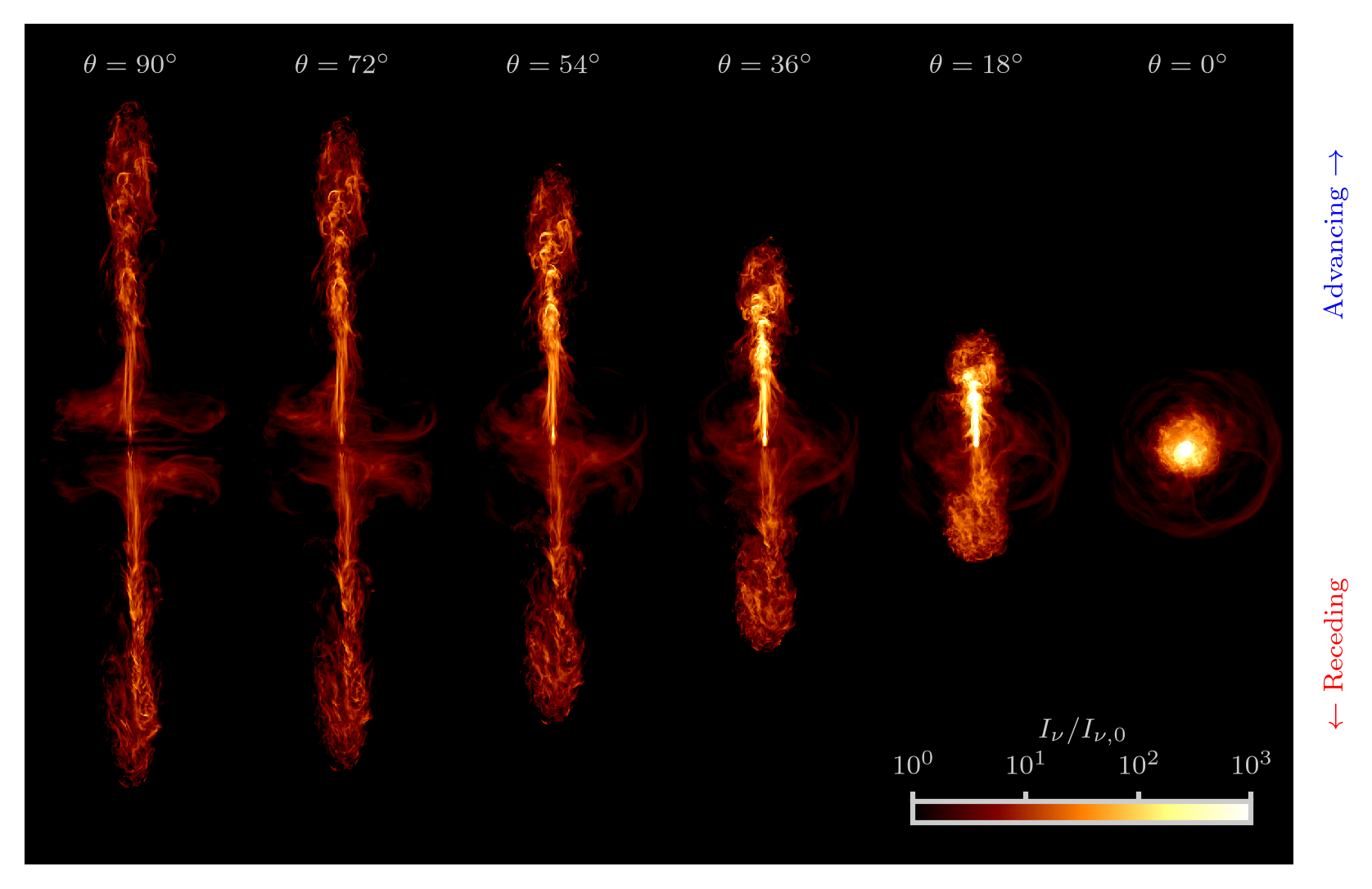}
    \caption{Synthetic radio observations generated by \texttt{cuDART} using data from a relativistic magneto-hydrodynamic simulation of a flickering AGN jet \citep{Elley_2026b}. Image data publicly available on \href{https://zenodo.org/records/20140021}{Zenodo}. Each panel depicts the same dataset (single snapshot) viewed from different orientations; due to relativistic beaming the jet is substantially brighter when the jet is closely aligned to the line-of-sight. See Section~\ref{sec:astrophysics} for a discussion of the observationally relevant relativistic and geometric effects recovered by \texttt{cuDART}.}
    \label{fig:agn_jets}
\end{figure*}

High-energy sources across the Universe produce relativistic outflows that are visible through their electromagnetic emission. Accreting black holes launch relativistic collimated outflows in the form of jets, driven by the extraction of rotational energy from a surrounding accretion disc \citep{Blandford_1982} or from the black hole ergosphere \citep{Blandford_1977}. Relativistic outflows are observable in stellar-mass black hole systems, such as X-ray binaries (XRBs, see \cite{Bahramian_2023} for review), but also in supermassive systems in the form of radio jets from Active Galactic Nuclei (AGN, see \cite{Hardcastle_2020} for review). There remain many uncertainties about the nature of these high energy systems, with researchers informing properties of both the outflow and the central engine from observations. 
Complex astrophysical environments frequently lack simple analytical descriptions, as such one method to learn more about these environments is from numerical studies. Relativistic (magneto-)hydrodynamic simulations give researchers information about the source structure that may not be resolvable in observations. By comparing the observational properties of simulations and real astrophysical systems, researchers are able to infer properties of the real system. For a single observation, a researcher may perform an array of simulations spanning the potential parameter space, using the result which best fits the observation to inform the properties of the source e.g. \cite{Savard_2025, Gasealahwe_2025}. 
\\~\\
In order for accurate comparison to be made between real astrophysical observations and numerical systems, simulation data must be rendered into synthetic observations. Studies that solve the radiative transfer equations readily yield electromagnetic observations by sampling the radiative flux on the simulation boundaries, but such simulations can prove both complex and expensive. Depending on the type of simulation used, assumptions may have to be made about emission mechanisms; for radio emission by synchrotron a common method is to assume equipartition and a power-law electron distribution to inform the local emissivity \citep{Longair_2011, Hardcastle_2013}. These methods only yield the spatial emissivity distribution, the researcher must then convert the 3D data into a 2D image. Once a 2D image has been formed, that image can be further processed by imaging pipelines to predict how such a source would be viewed through a real instrument e.g. \texttt{BLOB-RENDER} \citep{Savard_2025}. The most suitable method for computing the emergent intensity depends on the physics relevant to the system. In systems where attenuation and scattering are important, solving the radiative transfer equations as a post-processing step may be required; the code \texttt{Sirocco} performs this computation using Monte-Carlo sampling of photon packets \citep{Matthews_2025}. In optically thin environments, as is common for radio sources powered by synchrotron emission, computing the emergent intensity can be performed by simply integrating the emissivity along a line-of-sight. If the simulation data is discretised onto single homogenous resolution Cartesian mesh, then column summation can be used to compute emergent intensities along cardinal axes e.g. Figure 1 of \cite{Elley_2026}. This summation becomes substantially more complex for general orientations and simulation domains with heterogeneous resolution, as it is not known a priori what cells intersect with a given line-of-sight. Tackling this issue generally requires some form of ray-tracing, with methodology dependent on the simulation data structure. Many codes such as \texttt{RAMSES} \citep{Teyssier_2002} feature domains using adaptive mesh refinement, requiring the tracer to handle regions of disparate spatial resolution \citep{Barreira_2016}. The code \texttt{AREPO} adopts a Voronoi mesh structure \citep{Weinberger_2020}; raytracers developed for this code require careful handling of the irregular cell structure \citep{Nelson_2013}.
\\~\\
For relativistic sources there are additional complexities. The most complete imaging descriptions account for general relativistic effects, modelling the bending of light rays by strong gravitational fields \citep{White_2022}. Even for sources well described by flat spacetimes, there are special relativistic effects associated with the anisotropic beaming of radiation emitted by material moving close to the speed of light \citep{Savard_2025}. While oft overlooked, in systems which evolve rapidly with respect to their light crossing time, it is insufficient to consider only a single simulation snapshot when computing observations. Routines which account for the finite propagation speed of light are often referred to as ``slow-light'' models; methods of this type have recently been used to study jet structure close to the supermassive black hole in M87 \citep{Tsunetoe_2026}. Analytic forms exist for converting the rest frame emission from systems with homogenous velocity to the observer frame \citep{Lind_1985}, but no ready analytic form exists for general fields spatially distributed in the lab frame. See Section~\ref{sec:frames} for a discussion of the subtle difference between rest, lab and observer frames.
\\~\\
With \texttt{cuDART} (the CUDA/3DDDA accelerated ray tracing framework), we provide a means to form synthetic observations by calculating the emergent intensity from optically thin simulation data discretised onto Cartesian meshes in the lab frame. The code accepts as input either a single homogenous resolution mesh, or multiple heterogeneous sub-domains with locally homogeneous resolution, providing native support of simulations using mesh refinement. Using emissivities defined in the fluid rest frame, the code applies live relativistic beaming on a cell-by-cell basis allowing for observations from arbitrary viewing orientations. By scanning multiple simulation epochs for a single observation, the code is capable of accounting for the finite communication time between emitter and observer, allowing for the recovery of standard relativistic/geometric predictions without requiring additional assumptions about the property of the emitter (e.g. continuous vs discrete). In this paper, we show how rendering methods that do not consider light's finite communication time may produce synthetic observations that feature incorrect temporal evolution, morphologies and fluxes. All of the ray tracing calculations are performed on the GPU, allowing high-resolution synthetic observations to be rapidly computed from large simulation datasets. Earlier versions of the \texttt{cuDART} have already been used to produce synthetic observations of supernovae-jet interactions \citep{Gasealahwe_2025} and flickering AGN jets \citep{Elley_2026b}. Figure~\ref{fig:agn_jets} depicts synthetic radio observations made using \texttt{cuDART} with simulation data from \cite{Elley_2026b}.
This current paper acts as a companion to the release of \texttt{cuDART} as an open-source tool available to the astrophysics community, hosted on \href{https://github.com/hwhitehead/cuDART}{\texttt{GitHub}} and documented using \href{https://cudart.readthedocs.io/en/latest/index.html}{ReadTheDocs}. In Section~\ref{sec:algorithm} we discuss the underlying physics equations that must be solved to generate observations, and present the code's method for efficiently evaluating these equations using spatially and temporally discretised simulation data. In Section~\ref{sec:astrophysics}, we utilise a mock simulation dataset to verify the results of \texttt{cuDART} against the textbook predictions of relativistic and geometric theory. In Section~\ref{sec:code_structure}, we briefly discuss the code structure, documenting the runtime chronology and memory handling philosophy. Section~\ref{sec:performance_scaling} describes the main computational challenges for the render routine, documents how \texttt{cuDART} addresses each of these challenges and then reports performance metrics for different problem sizes and GPU architectures. In Section~\ref{sec:caveats} we discuss the limitations for the code, and avenues for future improvement. Finally, in Section~\ref{sec:conclusions} we conclude and summarise the paper.

\section{Algorithmic Implementation}
\label{sec:algorithm}
We describe the main calculation performed by the \texttt{cuDART} code, starting with an integral description and then embedding this integral within a simulation space discretised in space and time. We explain how this calculation is applied to the principle astrophysical use case of computing the emergent intensity from a relativistic system. 

\subsection{Path Integration}
\label{sec:path_integration}
\texttt{cuDART} is designed to operate on simulation data stored within a 3D Cartesian mesh defined in the lab frame. The lab frame is described by a static Minkowski spacetime, neglecting any cosmological effects, and is the standard basis for outputs from relativistic hydrodynamic codes. In its most general form, \texttt{cuDART} supports the calculation of some integrated line-of-sight function $F$, 
\begin{equation}
    \label{eq:general_integral}
    F(\bm{x}_0,\hat{\bm{s}},t_\mathrm{obs}) = \int_0^\infty f(\bm{x},\hat{\bm{s}},\bar{t}) \md s.
\end{equation}
Here, $\hat{\bm{s}}$ describes the orientation for a line-of-sight projected from an observer at origin $\bm{x}_0$, such that all spatial coordinates on the line-of-sight satisfy $\bm{x} = \bm{x}_0 + s\hat{\bm{s}}$. The traced field $f$ may be a function of space $\bm{x}$, time $t$ and viewing orientation $\hat{\bm{s}}$. The code currently only supports scalar $f$, though the method generalises readily to vector integration (e.g. for computing polarisation, see Section~\ref{sec:caveats}). Note that in Equation~\ref{eq:general_integral}, the local field $f$ is evaluated at $\bar{t}$ instead of $t_\mathrm{obs}$; this discrepancy arises due to finite time delay and defines the difference between lab and observer frames.

\subsection{Lab and Observer Frames}
\label{sec:frames}

In \texttt{cuDART} distinction is made between evaluating fields in the lab and observer frames. Alternative rendering codes often operate using the ``fast-light'' approximation, where an observation is formed using a single snapshot recorded at a homogeneous lab frame time. In reality, there will be a finite communication delay between an event occurring, and information of that event's existence reaching an observer. When considering radiation, the relevant event is the emission of a photon, which occurs earlier than the photon's reception by an observer due to a finite light travel time. The time delay between occurrence at $\bm{x}$ and observation at $\bm{x}_0$ can be expressed 
\begin{equation}
    \label{eq:t_delay}
    t_\mathrm{delay} = t_\mathrm{obs}-\bar{t} = \frac{|\bm{x} -\bm{x}_0|}{c} = \frac{s}{c},
\end{equation}
where $c$ is the communication speed (e.g. the speed of light, if tracing radiation). This finite communication time $t_\mathrm{delay}$ between a spatial location and the observer defines the difference between the lab frame and the observer frame. Both operate using fields defined at the same spatial location $\bm{x}$, but evaluate the fields at different times. When observing radiation, integrating in the lab frame would capture photons that were \textit{emitted} simultaneously. Conversely, integrating in the observer frame time would capture photons that are \textit{received} simultaneously, which is physically consistent with how real observations are generated. \texttt{cuDART} operates in the observer frame when using the ``lookback'' routine, which accounts for the finite time delay, allowing for the recovery of a range of relativistic and geometric phenomena (see Section~\ref{sec:astrophysics}). 
\\~\\
In summary, a snapshot at time $t_\mathrm{obs}$ in the lab frame describes a set of spacetime locations $\left(t_\mathrm{obs},\bm{x}\right)$ which are simultaneous in the limit of infinite communication speed. Conversely, a snapshot at $t=t_\mathrm{obs}$ in the observer frame defines a set of points in spacetime $\left(\bar{t}(\bm{x}),\bm{x}\right)$ which \textit{appear} simultaneous to the observer due to the finite communication speed between spatial position $\bm{x}$ and the observer $\bm{x}_0$. In the latter case, the local time $\bar{t}(\bm{x})$ that the fields are evaluated at is spatially inhomogeneous. Unlike the rest and lab frames, the observer frame is not transformed to using a Lorentz boost, but by a shear in time that is dependent on the observer-event separation. In systems where the communication crossing time is much shorter than the evolution timescale of the system, the observer frame is identical to the lab frame. This limit can be recovered by allowing the communication speed to diverge
\begin{equation}
    \lim_{c\to\infty}\left[\bar{t}(\bm{x})\right] = t_\mathrm{obs}\;\forall\;\bm{x}.
\end{equation}
In this fast-light limit, the local time is then spatially homogeneous, but this assumption does not hold for general astrophysical systems.

\subsection{Discretisation}

\label{sec:discretisation}
While the target calculation is integration along a line-of-sight, in practice, spatially/temporally distributed quantities returned by a simulation (such as the emissivity and velocity) will only be sampled at finite cadence, quantised spatially onto a mesh and temporally as snapshots at constant lab-frame time $t$. In Section~\ref{sec:dda}, we will show how the 3D digital differential analyser framework supports the accurate calculation of line weights for a line-of-sight embedded within a spatially meshed domain, allowing the integration to be discretised as 
\begin{equation}
    F(\bm{x}_0,\hat{\bm{s}},t_\mathrm{obs}) = \sum_n f_n(\bar{t}_n) \Delta s_n,
\end{equation}
where here the subscript $n = (i_n, j_n, k_n)$ labels the spatial indices for all cells that intersect with the line-of-sight. The weight factor $\Delta s_n$ describes the line element of the line-of-sight within cell $n$. Suppose additionally, that all fields are quantised in time, over snapshots indexed by $m$ at fixed cadence $\Delta t$ such that
\begin{equation}
    f_{n,m} = f_n(t_m), \quad t_m=m\Delta t.
\end{equation}
For $\bar{t}_n \neq m\Delta t$, sampling the field at $f_n(\bar{t}_n)$ requires interpolation between adjacent physical states. We introduce $\bar{m} \equiv  \left \lfloor \bar{t} / \Delta t \right \rfloor$ as the index of the ``early'' timestep and the $\delta_{n,m}(s,t)$ as the interpolation parameter between the local evaluation time $\bar{t}_n$ and the global discretised time $t_m$
\begin{equation}
    \delta_{n,m}(t_\mathrm{obs}) = \frac{|\bar{t}_n-t_m|}{\Delta t} = \frac{|t_\mathrm{obs}-\frac{s_n}{c}-m\Delta t|}{\Delta t}.
\end{equation}
We adopt a  linear interpolation kernel $W_{n,m} \in \left[0,1\right]$ expressed as 
\begin{equation}
    W_{n,m}(t_\mathrm{obs}) = 
    \begin{cases}
        1-\delta_{n,m}(t_\mathrm{obs}), &m \in [\bar{m}, \bar{m}+1] \\
        0, &m \notin [\bar{m}, \bar{m}+1].
    \end{cases}
\end{equation}
The weighting kernel allows for smooth transition between states $[f_{n,\bar{m}},f_{n,\bar{m}+1}]$ for $\bar{t}_n \in \left[\bar{m}\Delta t, (\bar{m}+1)\Delta t\right]$, such that for any field sampled at a time $\bar{t}_n$,
\begin{equation}
    f_n(\bar{t}_n) \equiv \sum_{m}W_{n,m}f_{n,m}= W_{n,\bar{m}} f_{n,\bar{m}} + W_{n,\bar{m}+1} f_{n,\bar{m}+1}.
\end{equation}
This weighting allows intermediary physical states to be approximated by summing over the temporally adjacent states. The full line-of-sight computation is then
\begin{equation}
    \label{eq:general_sum}
    F(\bm{x}_0,\hat{\bm{s}},t_\mathrm{obs}) = \sum_m \sum_n W_{n,m} f_{n,m} \Delta s_n,
\end{equation}
where the sum over $n$ describes a path summation through the domain (with line weighting $\Delta s_n$), and the sum over $m$ combines contributions from each simulation snapshot to the path summation (through the time/space dependent weighting $W_{n,m}$). If the the communication delay $t_\mathrm{delay}$ (see Equation~\ref{eq:t_delay}) is very short compared to the evolution of the simulation e.g. for  $c\to\infty$, then the local evaluation time will become homogeneous ($\bar{t}(\bm{x})=t_\mathrm{obs}\;\forall\;\bm{x}$). In this limit, 
\begin{equation}
    \lim_{c\to\infty}\left[W_{n,m}(t_\mathrm{obs})\right] = 
    \begin{cases}
        1, &m = \bar{m} \\
        0, &m \neq \bar{m}.
    \end{cases}
\end{equation}
Here the summation over $m$ can be neglected and only the snapshot closest in time to the observer will be used. This limit is the standard for many existing numerical visualisation schemes, in part due to the cost associated with storing and processing a large number of simulation snapshots. However only the more general weighting using multiple snapshots accurately recovers the full spectrum of relativistic and geometric behaviour as predicted by theory (see Section~\ref{sec:astrophysics}).

\subsection{3D Digital Differential Analyser}
\label{sec:dda}

\begin{figure*}
    \centering
    \includegraphics[width=2\columnwidth]{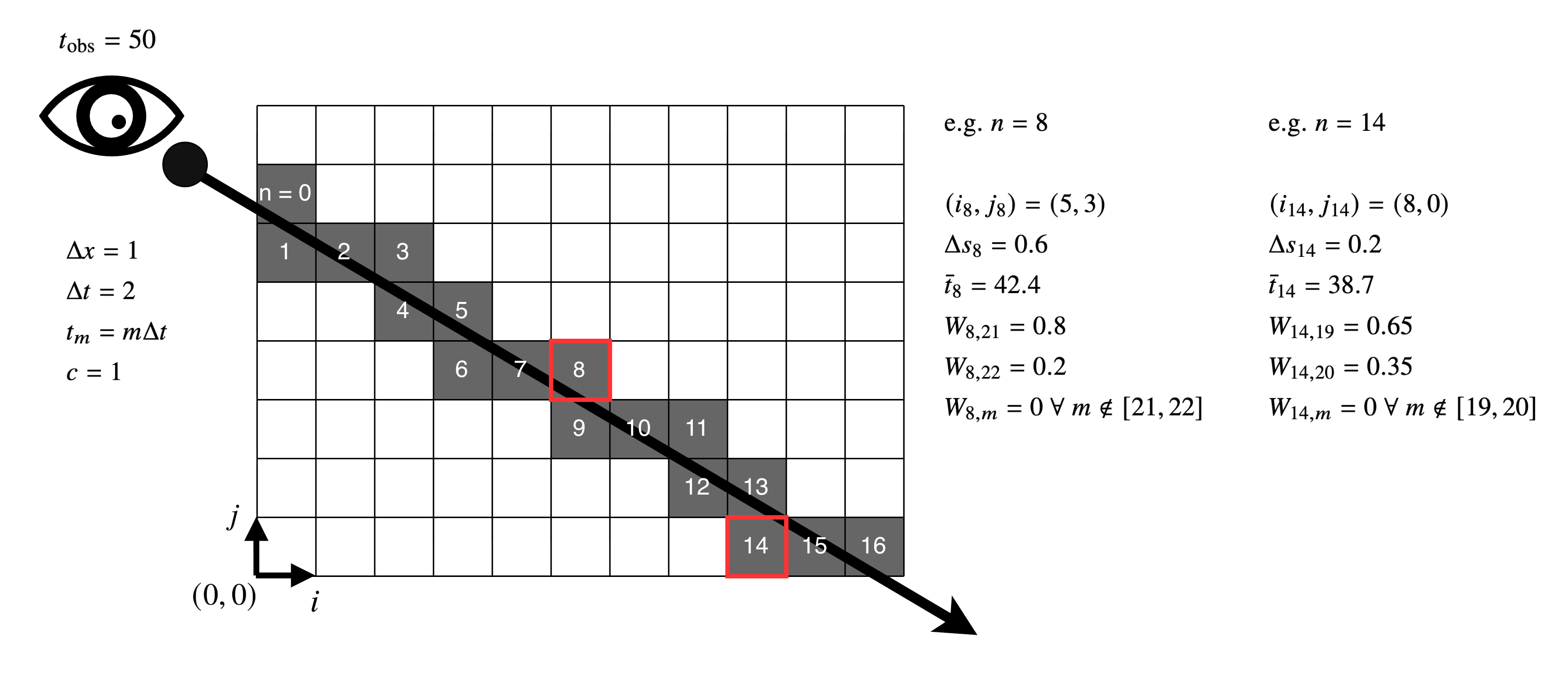}
    \caption{Schematic description of the ray-tracing process for a single observation made at $t_\mathrm{obs}$, indicating how the cell-by-cell contributions to the line-of-sight summation are computed by identifying the indices of all cells on the line-of-sight $n=(i_n,j_n)$, the line element within each cell $\Delta s_n$, the evaluation time $\bar{t}_n$ and the temporal weighting factors for adjacent simulation snapshots $W_{n,m}$. For this schematic system (with spatial resolution $\Delta x$, snapshot cadence $\Delta t$), these quantities are evaluated explicitly for cells $n=8$ and $n=14$ (listed on the right). We can see that despite being viewed at the same observer time $t_\mathrm{obs}$, cell 14 is sampled at an earlier time than cell 8 ($\bar{t}_{14}=38.7$ vs. $\bar{t}_8=42.4$), as the cell lies deeper in the domain. The code uses these cell-by-cell quantities to sum the field along the line-of-sight (see Equation~\ref{eq:general_sum}).}
    \label{fig:dda}
\end{figure*}

In order to perform a summation through a simulation domain along some general line-of-sight described by an origin $\bm{x}_0$ and orientation $\hat{\bm{s}}$, we need an efficient method to compute both the indices for all cells along the line of sight $n=\{i_n,j_n,k_n\}$, and local integration weights such as the line element $\Delta s_n$. If the line-of-sight is along a cardinal axis for a Cartesian mesh (e.g. $\hat{\bm{s}} \parallel\hat{\bm{z}}$), the intersecting cells are trivially all cells along a specific row/column. For more general sight lines, additional computation is required. To perform this identification we adapt an algorithm initially developed for ray tracing in computer graphics, the 3D Digital Differential Analyser \citep{Fujimoto_1986} (hence forth 3DDDA). This algorithm was designed to rapidly traverse bounding volumes containing a more complex sub-geometry to be rendered. For a scene containing a large number of polygons, it is much more efficient to divide the domain into a regular grid, calculate which grid cells a ray intersects with, and the only test ray-polygon intersections for objects within these specific grid cells. 
\\~\\
In brief, the action of 3DDDA can be described as an iterative propagation of a ray through a regular Cartesian (though not explicitly cubic) mesh. This propagation can performed rapidly by recognising that for a fixed ray orientation $\hat{\bm{s}}$, the separation between successive intersections with parallel cell faces is constant. The ray will cross cell faces with normals in the $\hat{\bm{x}}_i$ direction with line element separation
\begin{equation}
    \Delta s_i = \frac{\Delta x_i}{\hat{\bm{s}}\cdot\hat{\bm{x}}_i},
\end{equation}
where here $\Delta x_i$ is the (constant) cell size along the $\bm{x}_i$ axis. In 3D $i\in[0,1,2]$ (for $\bm{x},\bm{y}$ and $\bm{z}$), though the underlying algorithm generalises to arbitrary dimensions. If the cell in which the ray enters the mesh can be identified (usually by an inexpensive single cuboid-ray intersection test), then by making comparison between the distances to the next cell face intersection, the index of the next cell along the ray path can be identified. For a full description of the algorithm, see the original paper \citep{Fujimoto_1986}  and also this excellent guide on acceleration structures \citep{scratchapixel}. In the original 3DDDA algorithm, the code would then search each of these bounding volumes for polygons to intersect; in the context of \texttt{cuDART}, we have no underlying polygons to trace, but by treating simulation cells as bounding volumes we can identify all simulation cells intersecting a line of sight. We can compute other local quantities such as $\Delta s_n$ and $\bar{t}_n$ on the fly, allowing for exact computation of line integrals/summations in the form of Equation~\ref{eq:general_sum}. 
\\~\\
Figure~\ref{fig:dda} gives a schematic description of \texttt{cuDART}'s calculation of local contributions to path integration showing a single trace through a 2D mesh (the procedure readily generalises to higher dimensions). The algorithm identifies all cells $n$ along the line-of-sight, calculates the line element within each cell $\Delta s_n$, the local evaluation time $\bar{t}_n$ and the temporal weights $W_{n,m}$ for each simulation snapshot $m$. The figure compares the weights and relevant snapshots for two cells in the ray path, indexed as $n=8,14$. Cell $n=14$ features a grazing intersection with the ray, with a line element of only $\Delta s_{14}=0.2$ vs. $\Delta s_8=0.6$. As cell $n=14$ is deeper within the domain, it samples snapshots that are earlier in the simulation evolution ($m=19,20$ vs. $m=21,22$). The code performs these calculations for all cells along the line-of-sight, for all pixels in the image plane and for all images specified by the user, summing the result. Despite this complexity, path summation usually occupies only a small fraction of the total runtime; see Section~\ref{sec:performance_scaling}.

\subsection{Computing Intensity}
\label{sec:intensity_calculation}
Having defined the general operation of the line integral computation, we now apply it to \texttt{cuDART}'s principle use case: generating synthetic electromagnetic observations. We seek to evaluate the emergent monochromatic intensity from a simulation, defined as 
\begin{equation}
    I_\nu = \frac{\md E}{\md A \md t \md \nu \md\Omega},
\end{equation}
describing the energy $\md E$ incident on a surface $\md A$ from a solid angle $\md\Omega$, in time and frequency intervals $\md t$ and $\md\nu$. In an optically thin system, where the effects of attenuation and scattering can be neglected, the intensity can be calculated directly by path-integrating the emissivity.
\begin{equation}
    \label{eq:def_intensity_int}
    I_\nu(\bm{x}_0,\hat{\bm{s}},t_\mathrm{obs}) = \int_0^\infty j_\nu (\bm{x},\hat{\bm{s}},\bar{t}) \md s,
\end{equation}
where here $j_\nu$ is the monochromatic emissivity in the observer frame. This emissivity is a function of $\hat{\bm{s}}$, so may be anisotropic. For most astrophysical environments it is easier to compute the emissivity in the rest-frame of the emitter $S'$. Transformations between these frames are mapped by the Doppler factor, defined as
\begin{equation}
    D(\bm{x},\hat{\bm{s}},t) = \frac{1}{\Gamma(\bm{x},t) \left(1+\bm{\beta}(\bm{x},t)\cdot \hat{\bm{s}}\right)}, \quad 
    \Gamma(\bm{x},t) \equiv \frac{1}{\sqrt{1-\bm{\beta}(\bm{x},t)^2}},
\end{equation}
where $\Gamma(\bm{x},t)$ is the Lorentz factor and $\bm{\beta}(\bm{x},t) = \bm{v}(\bm{x},t)/c$ is the emitter velocity in units of $c$. Crucially, frequencies in each frame are related by $\nu = D\nu'$. The quantity $j_\nu/\nu^2$ is Lorentz invariant due to the competing effects of length contraction and time dilation, allowing the emissivity in each frame to be related as 
\begin{equation}
    j_\nu(\bm{x},t)= D(\bm{x},t)^2 j'_{\nu'}(\bm{x},t),
\end{equation}
where $\nu'=D^{-1}\nu$. Equation~\ref{eq:def_intensity_int} can then be rewritten as 
\begin{equation}
    \label{eq:non-power-law}
    I_\nu(\bm{x}_0,\hat{\bm{s}},t_\mathrm{obs}) =\int_0^\infty  D(\bm{x},\hat{\bm{s}},\bar{t})^2j'_{\nu'}\left( \bm{x},\bar{t}\right) ds,
\end{equation}
In \texttt{cuDART} v1.0, all rest-frame emission is modelled as isotropic (independent of $\hat{\bm{s}}$). While rest frame isotropy is a reasonable assumption for synchrotron emission within a tangled magnetic field, structured fields can impart anisotropy. This anisotropy is not accounted for in the current codebase, see Section~\ref{sec:caveats} for future development plans. The integral form of Equation~\ref{eq:non-power-law} is favourable as all orientation dependence has been shifted to the Doppler factor, which can be computed on-the-fly. However, the Doppler shifting of the emitted radiation means that computing an observation at a single lab-frame frequency $\nu$ now requires sampling many different rest-frame frequencies $\nu'(\bm{x})=D(\bm{x})^{-1}\nu$. Sampling a range of frequencies at runtime would require the code to read in frequency-dependent emissivity data, which introduces significant computational cost. This cost can be avoided if the rest-frame emission can be modelled as a power-law\footnote{Technically more general functional forms for $j'_{\nu'}=j'(\nu')$ are feasible, but only power-law emission is supported in \texttt{cuDART} v1.0.}, such that $j'_{\nu'}\propto \nu'^\alpha$. In this case all rest frame emission can be related to a reference value
\begin{equation}
    j'_{\nu'}(\bm{x}) = j'_{\nu_0'}(\bm{x}) \left(\frac{\nu'}{\nu_0'}\right)^\alpha = \left(\frac{\nu}{\nu_0'}\right)^\alpha D(\bm{x},\hat{\bm{s}},\bar{t})^{-\alpha} j'_{\nu_0'}(\bm{x}) , 
\end{equation}
where here $\nu'_0$ is some reference frequency in the emitter rest frame, and $\alpha$ is the slope for the rest frame emissivity power-law. Rest-frame power-law emission profiles are commonly assumed for synchrotron radio sources; in practice the emissivity in a simulation may be informed directly from tracking a non-thermal electron distribution \citep{Vaidya_2018,Mukherjee_2021, Dubey_2024,Elley_2026}, or by estimation via assumptions of equipartition and scaling with local pressure \citep{Longair_1994, Longair_2011, Hardcastle_2013}. For example, if emission is dominated by a population of electrons whose power-law distribution in energy has a slope $p=2$, then the resulting rest-frame emission will also be a power law with $\alpha=-0.6$ (this value is the default assumed by \texttt{cuDART}, but can be set manually by the user). The rest-frame power-law assumption is reasonable provided that the observer frequency is not redshifted beyond the power-law break frequency, predicted to occur in the GHz range \citep{Vijay_2020}. Applying this power-law form to Equation~\ref{eq:non-power-law} yields the full integral form implemented by \texttt{cuDART}: 
\begin{equation}   
    \label{eq:full_int}
    I_\nu(\bm{x}_0,\hat{\bm{s}},t_\mathrm{obs})= \left(\frac{\nu}{\nu'_0}\right)^\alpha\int_0^\infty D(\bm{x},\hat{\bm{s}},\bar{t})^{2-\alpha}j'_{\nu'_0}(\bm{x},\bar{t})ds.
\end{equation}
In this form, \texttt{cuDART} requires only three inputs:
\begin{itemize}
    \item The cell monochromatic rest-frame emissivity: $j'_{\nu'_0}(\bm{x},t)$ 
    \item The cell velocity in units of $c$: $\bm{\beta}(\bm{x},t)$
    \item The power-law index for the rest frame emissivity: $\alpha$
\end{itemize}
With these inputs, the emergent intensity at any time, and from any orientation can be computed. As this integral is embedded within a simulation domain that is discretised in space and time (see Section~\ref{sec:discretisation}),  \texttt{cuDART} computes the emergent intensity as 
\begin{equation}
    \label{eq:intensity_sum}
    I_\nu(\bm{x}_0, \hat{\bm{s}},t_\mathrm{obs}) = \left(\frac{\nu}{\nu'_0}\right)^\alpha\sum_m \sum_n W_{n,m}D_{n,m}^{2-\alpha}j'_{\nu_0',n,m} \Delta s_n.
\end{equation}
A useful consequence of modelling the rest-frame emission as a power-law in frequency is that a single monochromatic intensity calculation contains all the relevant information for other observer frequencies e.g.
\begin{equation}
    I_{\nu_2}(\bm{x}_0,\hat{\bm{s}},t_\mathrm{obs}) = \left(\frac{\nu_2}{\nu_1}\right)^\alpha I_{\nu_1}(\bm{x}_0,\hat{\bm{s}},t_\mathrm{obs}) \quad \forall \; \nu_1, \nu_2.
\end{equation}
If the underlying rest-frame emission obeys a power-law with a spatially homogenous and static slope $\alpha$, the resultant intensity retains this power-law scaling. 

\section{Astrophysics Tests}
\label{sec:astrophysics}
\begin{figure*}
    \centering
    \includegraphics[width=2\columnwidth]{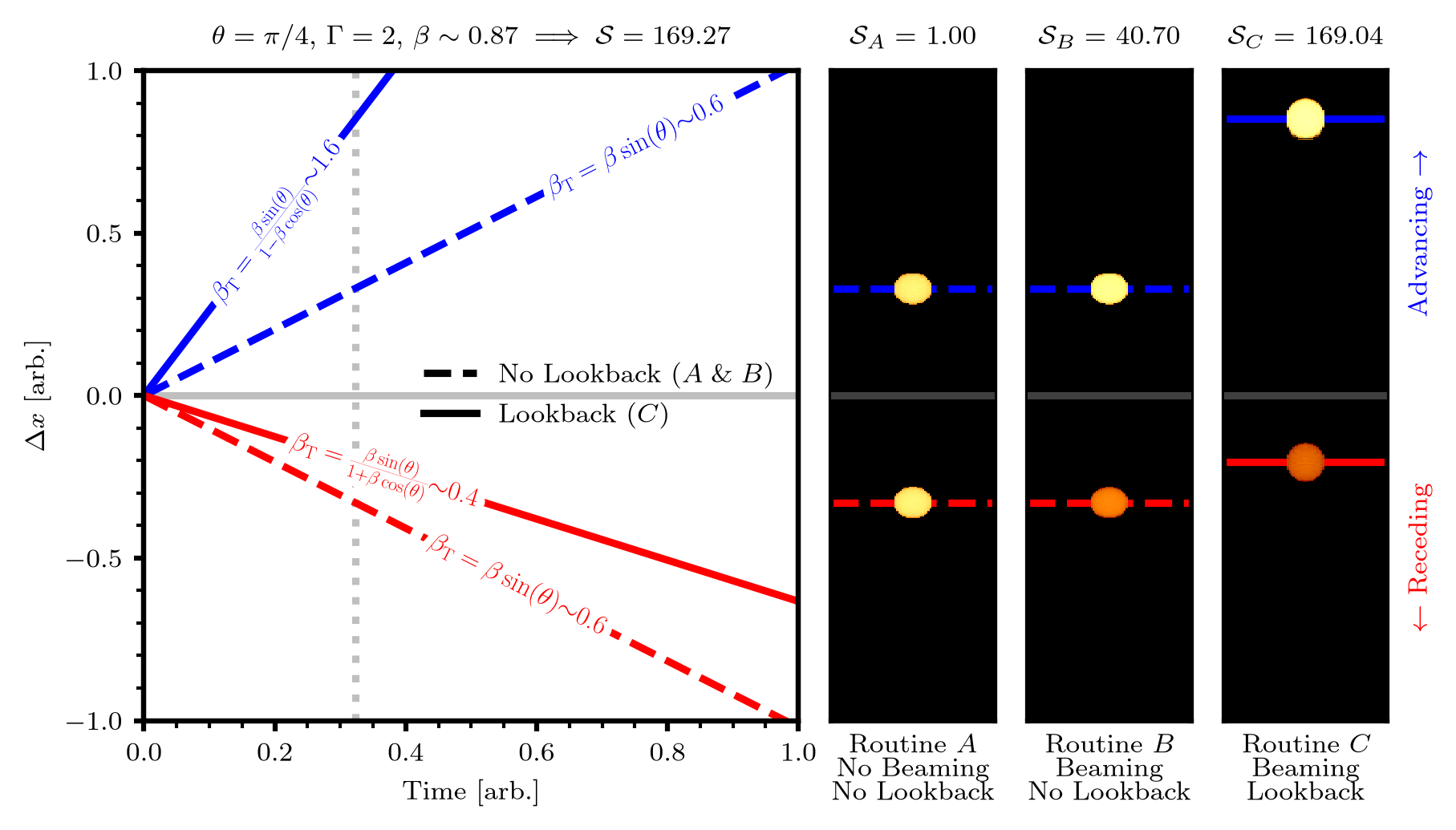}
    \caption{Comparison of observed properties for a mock simulation dataset hosting twin anti-parallel relativistic ejecta, when imaged using three different routines with increasing levels of complexity ($A$, $B$ and $C$). Without lookback ($A$, $B$), rendering is performed using a single simulation snapshot. With lookback ($C$), multiple snapshots are scanned to account for a finite communication time between source and observer.
    In the left panel, the observed transverse motion for systems with/without lookback. In the right panels, we show the synthetic observations generated under different routines at an observer time given by the grey dashed line in the left panel. Without beaming ($A$), the advancing and receding ejecta have the same brightness; when beaming is included ($B$, $C$) the advancing ejecta is substantially brighter (shown by the flux ratio $\mathcal{S}\equiv S^\mathrm{adv}_\nu/S^\mathrm{rec}_\nu$). Without lookback ($A$, $B$), the ejecta exhibit symmetric transverse motion ($\beta_\mathrm{T}\sim0.6$) and are imaged as oblate spheroids, coherent with their lab-frame morphology. When lookback is included ($C$), the proper asymmetric transverse motion is captured, with the advancing ejectum appearing to move faster ($\beta_\mathrm{T}\sim 1.6$) than the receding ejectum ($\beta_\mathrm{T}\sim0.4$). Further, the ejecta
    are observed as spheres, consistent with the relativistic/geometric predictions of the Penrose-Terrell effect. When lookback is enabled, the flux ratio between ejecta matches the expected theoretical value to $0.2\%$. As such, only routine $C$ produces synthetic observations that match the predicted transverse motion, morphology and flux; see Sections~\ref{sec:beaming}-\ref{sec:luminosity}.}
    \label{fig:triple_comp}
\end{figure*}

In this section we show how \texttt{cuDART} recovers standard astrophysical results as predicted by relativistic and geometric theory. As a demonstration environment, we analyse a mock simulation data set featuring twin anti-parallel ejecta travelling with bulk Lorentz factors of $\Gamma = 2$ ($v\sim0.87c$). These ejecta are modelled as spheres in their own co-moving frame and hence in the lab frame are oblate spheroids with an axial ratio of $1/\Gamma$ parallel to their motion (due to Lorentz contraction). For the purpose of this test, the emission within each ejectum is modelled as homogenous. The simple geometry of this mock data makes for easy comparison between synthetic observations and analytical expectations; more complex environments generally lack analytical comparisons. Mock data suites with this geometry (and others) can be built using tools provided in the \texttt{cuDART} regression suite.
\\~\\
Figure~\ref{fig:triple_comp} depicts, in its right-hand panels, images of the mock simulation data taken using three routines:
\begin{enumerate}[leftmargin=0.5cm, label=(\Alph*)]
    \item Not including relativistic beaming or lookback
    \item Including relativistic beaming, but not lookback
    \item Including relativistic beaming and lookback 
\end{enumerate}
In this section we will show that generally only this last treatment accurately recovers the observable properties (transverse motion, morphology and flux) predicted by relativistic and geometric theory. 

\subsection{Relativistic Beaming}
\label{sec:beaming}
The left two images of Figure~\ref{fig:triple_comp} (routines $A$ and $B$) compare renders made without and with relativistic beaming repsectively. In their own rest frames, the advancing and receding are identical, so when beaming is neglected the two ejecta exhibit the same brightness. Beaming introduces anisotropy to the lab frame emissivity, breaking this symmetry and resulting in an advancing ejectum that is significantly brighter. Similarly, the receding ejecta is dimmer than the case without beaming, as the emission is beamed away from the observer. Comparing the total flux ($S_\nu \propto \int I_\nu \md A$) emitted by the advancing and receding ejecta in the beamed case gives a flux ratio of $\mathcal{S}\equiv S^\mathrm{adv}_\nu/S^\mathrm{rec}_\nu\sim40$, set by the $D^{2-\alpha}$ scaling that enters into the intensity integral of Equation~\ref{eq:full_int}. This is actually still the \textit{incorrect} flux ratio, the true value is only recovered when lookback is also included, see Section~\ref{sec:luminosity} for discussion.

\subsection{Transverse Motion}
\label{sec:transverse_motion}
A textbook consequence of the finite travel time of light is a discrepancy between the true and observed transverse motion for a relativistic emitter. This is most obvious when the true velocity of the region is directed partially toward the observer; because the distance between emitter and observer is decreasing, the observed transverse motion is larger than reality. For an object travelling at $\beta=v/c$ at an inclination of $\theta$ to the observer's line-of-sight, the apparent transverse velocity of the object $\beta_T$ takes the form
\begin{equation}
    \beta_T = \frac{\beta\sin\left(\theta\right)}{1-\beta\cos\left(\theta\right)}.
\end{equation}
While the true velocity is constrained to $\beta \in [0,1]$ by relativity, the observer velocity is extremised with respect to $\theta$ at $\theta_\mathrm{crit}=\cos^{-1}(\beta)$; at this orientation $\beta_\mathrm{T}=\Gamma \beta$. Hence, for $\beta > 1/\sqrt{2}$, there exist orientations $\theta \sim \theta_\mathrm{crit}$ where $\beta_\mathrm{T} > 1$. In this scenario, the object appears to be moving faster than the speed of light (termed ``superluminal motion''). This result is \textit{only} recoverable when a finite light time delay is accounted for, hence synthetic observations which assume instantaneous communication between source and observer fail to report the proper transverse motion. The left panel of Figure~\ref{fig:triple_comp} shows the observed displacement of twin-ejecta moving at fixed velocity, with the right panels comparing renders made with ($C$) and without lookback ($A$ and $B$). Without lookback, both ejecta are observed to have the same transverse speed (dashed lines in left panel), but with lookback they exhibit the proper asymmetric motion with the advancing ejecta appearing to travel faster than the receding (solid lines in left panel). 

\subsection{Morphology}
\label{sec:morphology}

\begin{figure}
    \centering
    \includegraphics[width=\columnwidth]{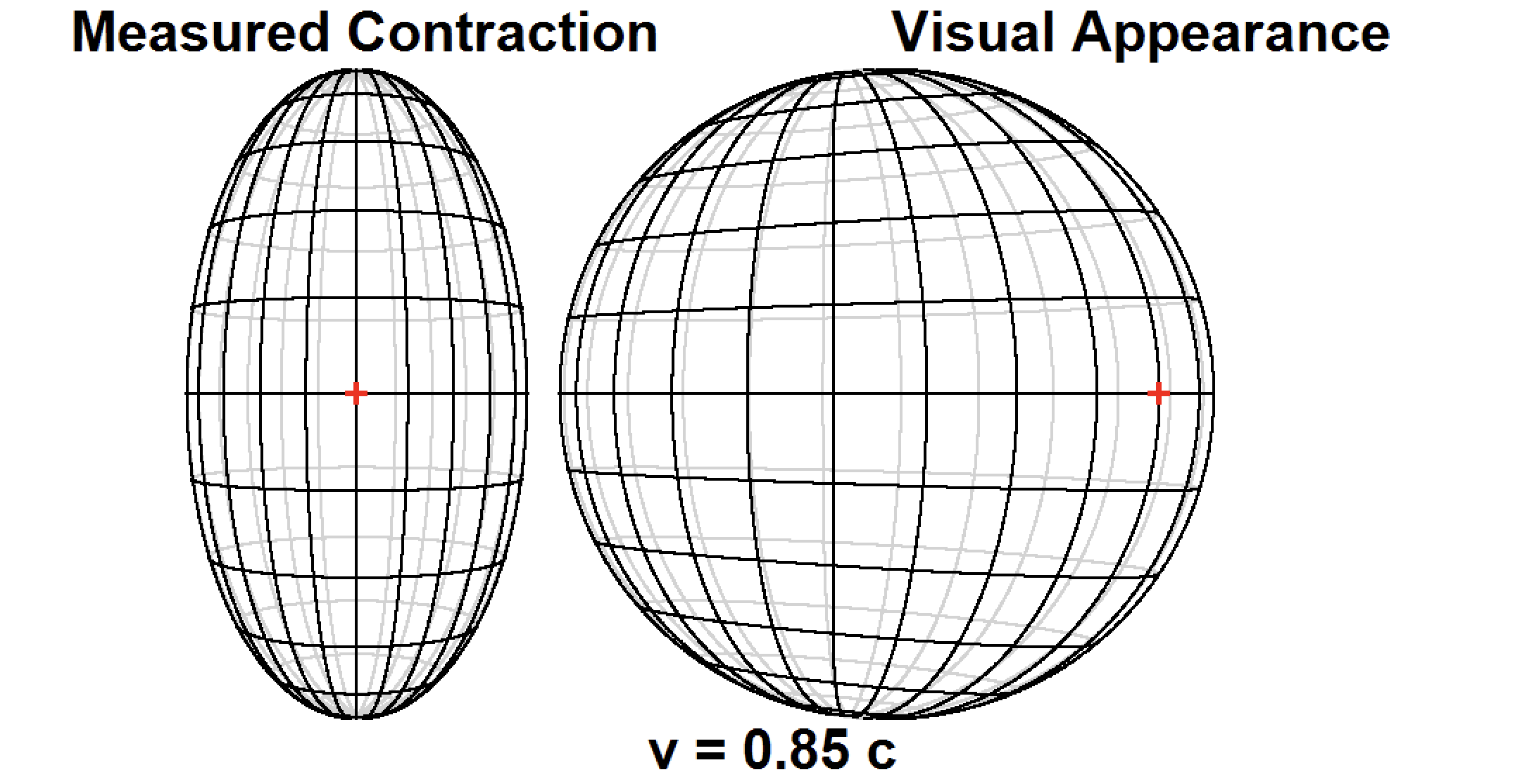}
    \caption{Schematic depicting the difference in the measured contraction and visual appearance for an object that is spherical in its rest frame. The closest point on the sphere's surface (marked with a red cross) appears rotated toward the direction of motion (right to left) light from this point arrives earlier than the rest of the sphere. This effect is known as Penrose-Terrell rotation. Figure adapted from animation by \href{https://commons.wikimedia.org/w/index.php?curid=125074651}{Wikipedia Commons}.}
    \label{fig:penrose_schematic}
\end{figure}

As discussed in Section~\ref{sec:transverse_motion}, allowing for a finite light travel time between emitter and observer can result in different observed motions. Similarly, the difference in light travel time between the near and far surfaces of an emitting region can result in morphological differences between the emitter structure as measured in the lab frame and as observed. This delay introduces an observable deformation in the emitter's geometry along its direction of motion, as the far surface is observed earlier in the object's motion than the near surface. As first discussed by \cite{Penrose_1959} and \cite{Terrell_1959} (and hence known as the Penrose-Terrell effect), this deformation opposes the size change imparted by Lorentz contraction, resulting in the observed size of the region matching the measurement made in the emitter's rest frame. In the scenario discussed by Penrose and Terrell, an emitter that is spherical in its own rest frame, while Lorentz contracted in the lab frame, is \textit{observed} to be spherical due to the differential lag time between near and far surfaces of the sphere. An image of the sphere would appear to be rotated toward the direction of motion: in the limit of $\beta \rightarrow 1$, the closest point on the sphere would appear to be the most displaced along the sphere's direction of motion. This rotation is invisible for the homogenous emitter used in this mock data set, see Figure~\ref{fig:penrose_schematic} for a schematic example of this effect. The right two panels of Figures~\ref{fig:triple_comp} ($A$ and $B$) make clear the importance of accounting for this lag time: failing to include lookback results in an image depicting (incorrectly), an elliptical emitter due to the lab-frame oblate spheroid structure. When lookback is included ($C$), the proper circular observation is recovered\footnote{Up to a precision set by the cadence of the simulation sampling, see Section~\ref{sec:aliasing} for discussion.}, as predicted by Penrose and Terrell. 

\subsection{Flux}
\label{sec:luminosity}
The subplot titles of Figure~\ref{fig:triple_comp} also compare the ratio of flux between advancing and receding ejecta. A standard result for discrete emission is that the rest-frame flux and observer flux for a source of homogeneous velocity are related by
\begin{equation}
    S_\nu = \int I_\nu d\Omega = \frac{D^{3-\alpha}}{L^2}\int j'_{\nu'}dV' \propto D^{3-\alpha},
\end{equation}
where we have used the Lorentz invariance of $I_\nu/\nu^3$  and  $d\Omega=dA'/L^2$ (the solid angle for a source at a distance $L$ to the observer): see \cite{Lind_1985} for a more detailed derivation. The ejecta travelling towards/away from the observer have identical structure in their own rest frames (labelled with primes), so the ratio of fluxes between advancing and receding ejecta follows as
\begin{equation}
    \mathcal{S}\equiv\frac{S_\nu^\mathrm{adv}}{S_\nu^\mathrm{rec}} = \left(\frac{D^\mathrm{adv}}{D^\mathrm{rec}} \right)^{3-\alpha} = \left(\frac{1+\beta \cos(\theta)}{1-\beta \cos(\theta)}\right)^{3-\alpha}.
\end{equation}
Figure~\ref{fig:triple_comp} computes this ratio by integrating over the pixels for the advancing and receding ejecta, for each of the three rendering routines. For $\theta=\pi/4$, $\Gamma=2$, the expected flux ratio is $\mathcal{S}=169.270$; the flux ratios for the three render routines are $\mathcal{S}_A=1.00$, $\mathcal{S}_B=40.70$ and $\mathcal{S}_C=169.04$. It is clear that case $A$  completely fails to capture the flux asymmetry, as no relativistic boosting has been applied. However even in case $B$, where boosting is included, the flux ratio is still incorrect. While on a cell-by-cell basis the emissivity has been properly boosted into the observer frame, by failing to track the proper emission morphology the total emergent flux has also been miscalculated. In contrast, when beaming and lookback is included (case $C$), the ratio of fluxes matches the theoretical result to within $0.2\%$. See Table~\ref{tab:method_comp} for a comparison between the observables yielded by each render routine and the theoretical predictions.
\begin{table*}
    \centering
    \begin{tabular}{c|c|c|c|c|c}
    \hline
         Method & Relativistic Beaming & Lookback & Transverse Velocity $\beta_\mathrm{T}$ & Morphology & Flux Ratio $\mathcal{S}$\\ \hline 
         $A$ & \xmark & \xmark & $\pm\beta \sin(\theta)$ & Ellipse & 1.0 \\  
         $B$ & \cmark & \xmark & $\pm\beta \sin(\theta)$ & Ellipse & 40.70 \\ 
         $C$ & \cmark & \cmark & $\frac{\pm\beta\sin(\theta)}{1\mp \beta\cos(\theta)}$ & Circle & 169.04 \\ \hline
         Theory & - & - & $\frac{\pm\beta\sin(\theta)}{1\mp \beta\cos(\theta)}$ & Circle & 169.27  \\ 
         \hline
    \end{tabular}
    \caption{Comparison of phenomena recovered by applying different rendering routines (labelled $A$, $B$ and $C$) to the twin-ejecta mock dataset (see Figure~\ref{fig:triple_comp}). Considered observables are the ejecta's transverse velocity, morphology and flux ratio (see Sections~\ref{sec:transverse_motion}-\ref{sec:luminosity}). Failing to account for both relativistic boosting and a finite time delay (lookback), results in observables which do not agree with the theoretical predictions. Only the most advanced routine $C$, \texttt{cuDART}'s default implementation, reproduces the theoretical predictions, recovering the flux ratio between advancing and receding ejecta to within $0.2\%$.}
    \label{tab:method_comp}
\end{table*}
Caution is warranted when attempting to reproduce this effect using a single simulation snapshot. The formulation of \cite{Lind_1985} makes clear the need for using $D^{3-\alpha}$ Doppler factors when boosting from the fluid rest frame to the lab, but this only holds if integration is performed in the fluid \textit{rest} frame. For a source of homogenous velocity, integrating with $D^{3-\alpha}$ Doppler factors in the \textit{lab} frame introduces an erroneous overestimation for the emitter volume (and hence emergent flux) by a factor $\Gamma$. If applied in the lab frame, the formula of \cite{Lind_1985} would yield incorrect absolute flux measurements for both the receding and advancing emitters, but would maintain a correct flux ratio due to a common erroneous multiplicative factor. For less idealised astrophysical systems, in which the velocity is unlikely to be homogeneous or static in time, there will not exist a global Lorentz/Doppler factor to convert between fluid and observer frames, requiring a numerical scheme to resolve.

\subsection{Why Lookback Matters}
\label{sec:lookback_matters}
It should be clear from Sections~\ref{sec:transverse_motion}-\ref{sec:luminosity} that while boosting the rest-frame emissivity to the lab frame is a requirement for imaging, alone it is insufficient to recover the full relativistic and geometric observational predictions. Only by including this beaming \textit{and} accounting for a finite communication time between an emitting region and the observer can accurate synthetic observations be formed. The importance of this effect is interpreting observations of relativistic sources is well known, for example in motivating the asymmetry in discrete ejecta from XRBs \citep{Espinasse_2020} and observability of advancing/receding jets at different epochs \citep{Stephens_2026, Cooper_2026}. While some recent numerical studies have included this effect to accurately recover the observed trajectories of relativistic ejecta \citep[e.g.][]{Savard_2025}, this effect is generally not included as part of existing imaging pipelines, preventing the generation of self-consistent synthetic observations. The accounting of a finite communication time, termed ``lookback'' in the \texttt{cuDART} framework, is included by default, requiring the user to provide a series of simulation snapshots in time instead of at a single epoch.
\\~\\
The toy model used to demonstrate these discrepancies features emitting regions which are static in their own rest frames (the emissivity of each region does not change, their velocity is constant and their shape unchanged). In using this simple toy model, we can make direct comparison to known theoretical results for the expected motion, morphology and fluxes. In a less idealised astrophysical setting, none of these static properties are assured and there may exist no tractable analytical expectations. Such cases require numerical calculation to generate accurate observations. 
\\~\\
It is important to note that in some systems it is reasonable to ignore the finite speed of light. If the morphology of a source (as defined in the lab frame), evolves slowly compared to its light self-crossing time, then the communication time between source and observer can be treated as effectively instantaneous and rendering can be performed on a snapshot-by-snapshot basis. In the language of \cite{Lind_1985}, this is equivalent to treating the lab-frame as the ``pattern frame''. This assumption produces reasonable fluxes for large-scale AGN jets as the advance speeds of jets into the circum-galactic medium is usually much slower than the speed of light; see \cite{Elley_2026b} for renderings of such a system using an earlier version of \texttt{cuDART}. However, caution is warranted when visualising rapidly evolving structures, such as knots in the jet beam, or comparing between the advancing and receding jets at late times as here the light time delay can be comparable to the dynamical time. Recovering the full spectrum of observable properties requires a scheme capably of accounting for a finite speed of light: these effects are accounted for by default in v1.0 of \texttt{cuDART}.

\section{Code Structure}
\label{sec:code_structure}
The computationally-intensive portion of \texttt{cuDART} is written in C++/CUDA, with domain traversal, beaming, temporal interpolation and summation all performed on the GPU. The user can interact with the C++ API via a frontend written in Python, which provides a simple class structure to support the labelling of simulation data, calling the C++ render routine, and processing the output files into human-readable figures (e.g. using \texttt{matplotlib}). A full description of the C++ and Python API can be found on \href{https://cudart.readthedocs.io/en/latest/index.html}{ReadTheDocs}. Given the efficiency of the actual render call, \texttt{cuDART} attempts to make as few expensive file read and memory copy operations as possible. To facilitate the reading/writing of \texttt{.npy} files using C++, we bundle the \texttt{libnpy} library (available publicly on \href{https://github.com/llohse/libnpy}{GitHub}) within the \texttt{cuDART} codebase. The simulation data read by the renderer may be composed either of a single mesh of homogeneous resolution, or multiple sub-domains of globally heterogenous, but locally homogeneous resolution; within \texttt{cuDART} these two operational forms are termed ``unlabelled'' and ``labelled'' respectively. Labelled mode allows \texttt{cuDART} to render simulation data yielded by codes utilising mesh refinement without the need for an intermediate regularisation step. More about the labelled/unlabelled running modes can be found in the \href{https://cudart.readthedocs.io/en/latest/inputs.html}{documentation}.
\\~\\
The execution chronology varies significantly depending on whether the user is running with or without the lookback routine; see Section~\ref{sec:algorithm} for implementation and Section~\ref{sec:astrophysics} for the implications of this routine.

\begin{figure*}
    \centering
    \includegraphics[width=2\columnwidth]{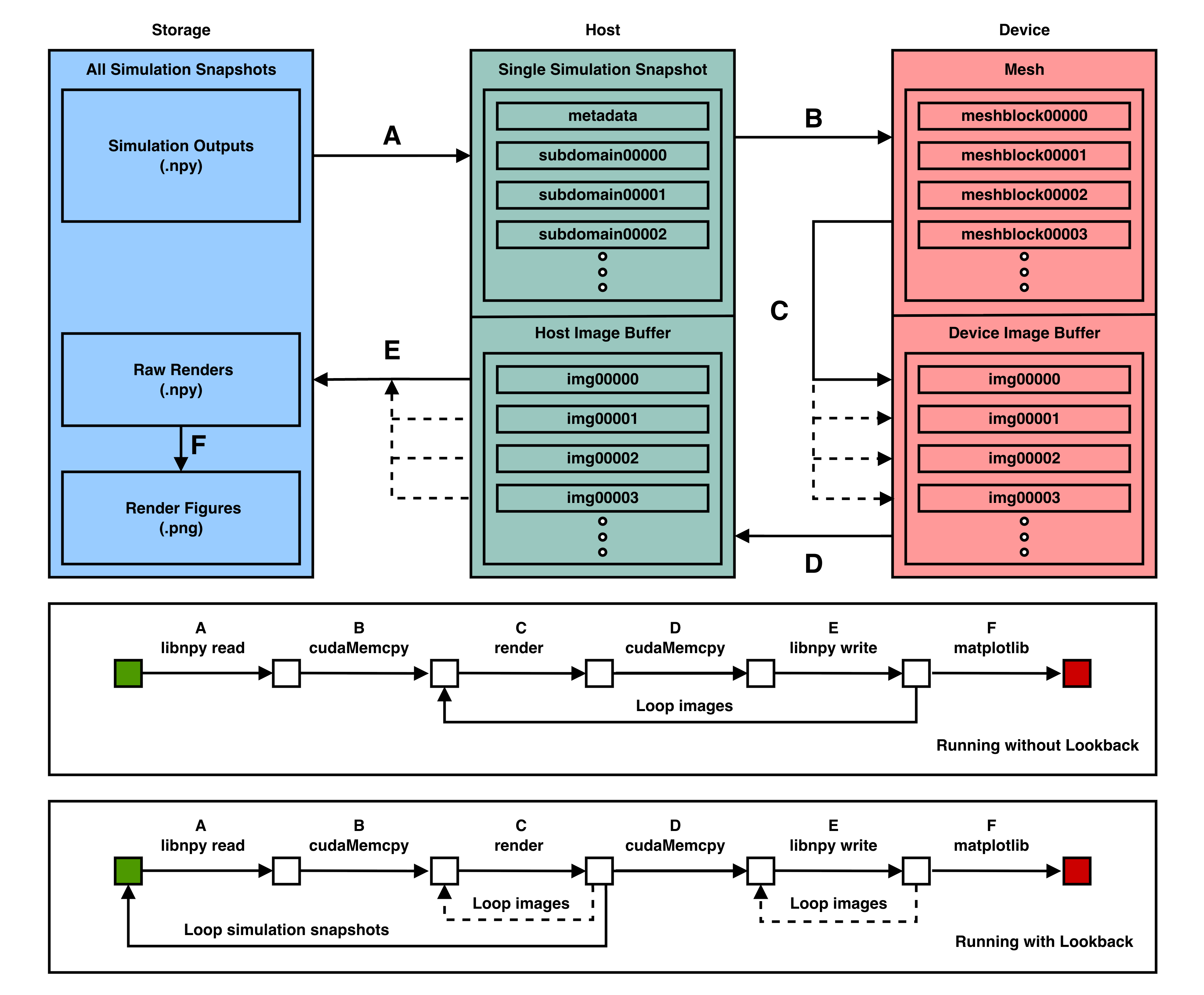}
    \caption{Schematic description of the runtime chronology, for renders performed both with and without the lookback routine. Data is loaded from storage into host memory using \texttt{libnpy}, and then copied to the device. Without lookback, images are rendered, copied and saved to storage within loop. With lookback, partial images are written additively to a communal buffer and the complete images are only copied and saved once all snapshots have been looped over. This minimises the number of  read/write/transfer operations at the expense of maintaining a persistent buffer for all images in device memory. The raw image files can be converted into human-readable figures using the Python frontend and \texttt{matplotlib}.}
    \label{fig:structure}
\end{figure*}

\section{Performance and Scaling}
\label{sec:performance_scaling}
In this section we discuss briefly the computational complexity posed by attempting to simultaneously raytrace multiple snapshots of large simulation datasets and explain how \texttt{cuDART} addresses this complexity. We restrict our discussion to the more computationally expensive lookback routine. Performance is tested on a mock simulation dataset with a domain composed of $D^3$ cells, sampled over $M$ snapshots. Rendering is then performed using $N$ cameras, each with $P^2$ pixels. The mock dataset is composed of a single homogeneous mesh, with the code operating in ``unlabelled'' mode; running in ``labelled'' mode makes the render operation more expensive by requiring an additional loop over sub-domains, but the effect is usually sub-dominant to file read bottlenecks. Generally, the main contributors to the \texttt{cuDART} runtime can be categorised as 
\begin{enumerate}[leftmargin=0.5cm, label=(\Alph*)]
    \item Render Operations: summing along ray paths 
    \item Memory Operations: copying data between host (RAM, CPU memory) and device (VRAM, GPU memory)
    \item File Operations: reading data from storage into  memory
\end{enumerate}
In Section~\ref{sec:render_optimisation}, we discuss how \texttt{cuDART} optimises the first step to the point that the runtime is entirely bottlenecked by memory/file operations (Section~\ref{sec:memory_optimisation}). Section~\ref{sec:performance} then shows how runtime scales with the problem size for a range of GPU architectures, exhibiting how in the file-read-bottlenecked regime performance is largely agnostic of architecture. 

\subsection{Render Operations}
\label{sec:render_optimisation}
In the naive scenario, where intersection must be tested between every pixel ray and every cell, across each camera and snapshot, the number of intersection tests grows rapidly with the problem size, scaling as $O(D^3MNP^2)$. For a relatively modest input parameter set of $\{D,M,N,P\}=\{512,100,100,512\}$, a brute force calculation requires a staggering $10^{17}$ intersection tests. The vast majority of these tests yield no intersection, resulting in a large number of wasted computations. The total number of relevant computations scales as $O(DNP^2)$, in the example above $10^{10}$ intersections between pixel rays and simulation cells must be considered, substantially fewer than the brute force estimate, but not insignificant. \texttt{cuDART} applies multiple methods to massively reduce the cost associated with this intersection and path summation calculation:
\begin{itemize}
    \item \textit{3DDDA}: only consider cells in ray path
    \item \textit{Fast-forward}: skip cells on ray path with no temporal overlap
    \item \textit{GPU parallelisation:} compute pixel values simultaneously
\end{itemize}
The 3DDDA algorithm allows the renderer to track a ray through a Cartesian domain, considering only cells on the ray path instead of the full domain (see Section~\ref{sec:dda}). This reduces the complexity of the path summation for a singe pixel from  $O(D^3M)\rightarrow O(DM)$, massively reducing the computational cost. With lookback enabled, because only cells with a temporal overlap with the camera contribute to the path summation, many of the weights $W_{n,m}$ on the ray path will be zero. The fast-forward routine ensures that the ray path is only traversed for the portion where the weighting is non-zero, this further reduces the per-pixel complexity scaling from $O(DM)$ to simply $O(D)$. 
Further, by deploying the render on the GPU, multiple pixel values can be computed simultaneously on separate threads. Because the problem requires no cross-talk between threads \citep[the problem is intrinsically ``embarrassingly parallel'', as coined by][]{Moler_1986}, this vectorisation scales exceptionally well. Depending on the number of concurrent threads available on the GPU, the code may be able to compute all pixels in a single wave, potentially removing all render scaling with $P$.

\subsection{Memory and File Operations}
\label{sec:memory_optimisation}
One bottleneck that is harder to avoid is the cost associated with reading simulation data from storage, and copying this data onto the device. Due to optimisation in the actual render routine (see Section~\ref{sec:render_optimisation}), these read/copy costs usually occupy a significant fraction of the runtime, largely dependent on the quality of the file storage network rather than the device hardware. To address this, \texttt{cuDART} attempts to make as few read and copy operations as possible, reading each simulation snapshot only once and opting to store image data in composite buffers instead of making partial writes to storage (see Figure~\ref{fig:structure}). Future optimisation routines are currently in the testing phase, including ``flexload'', which will avoid loading snapshots or calling render kernels with no temporal overlap. 
\\~\\
While the cost of both file read and memory copy operations scale as $O(MD^3)$, memory throughput is generally much higher than file read/write speeds. When file read speeds bottleneck performance, savings could be made by parallelising the file read process, using libraries such as HDF5 \citep{The_HDF_Group_Hierarchical_Data_Format}. An alternative mitigation would be to load the only the relevant sections of each simulation snapshot, which would likely require partitioning the domain into smaller bounding volumes (which has its own read/write costs) and then running intersection tests before loading the data. Implementing such a routine within \texttt{cuDART} is feasible, but not included in the v1.0 release, in part because this only reduces runtime by including an expensive pre-render step. File read bottlenecks make portable commentary on performance difficult, as the runtime will likely depend on the user's local storage environment. We recommend users perform their own profiling, supported by tools provided within \texttt{cuDART} that invoke and summarise results fron NVIDIA's Nsight Systems profiling toolkit (see \href{https://cudart.readthedocs.io/en/latest/performance.html}{ReadTheDocs} for recommended usage).

\subsection{Performance Metrics}
\label{sec:performance}

\begin{figure*}
    \centering
    \includegraphics[width=2\columnwidth]{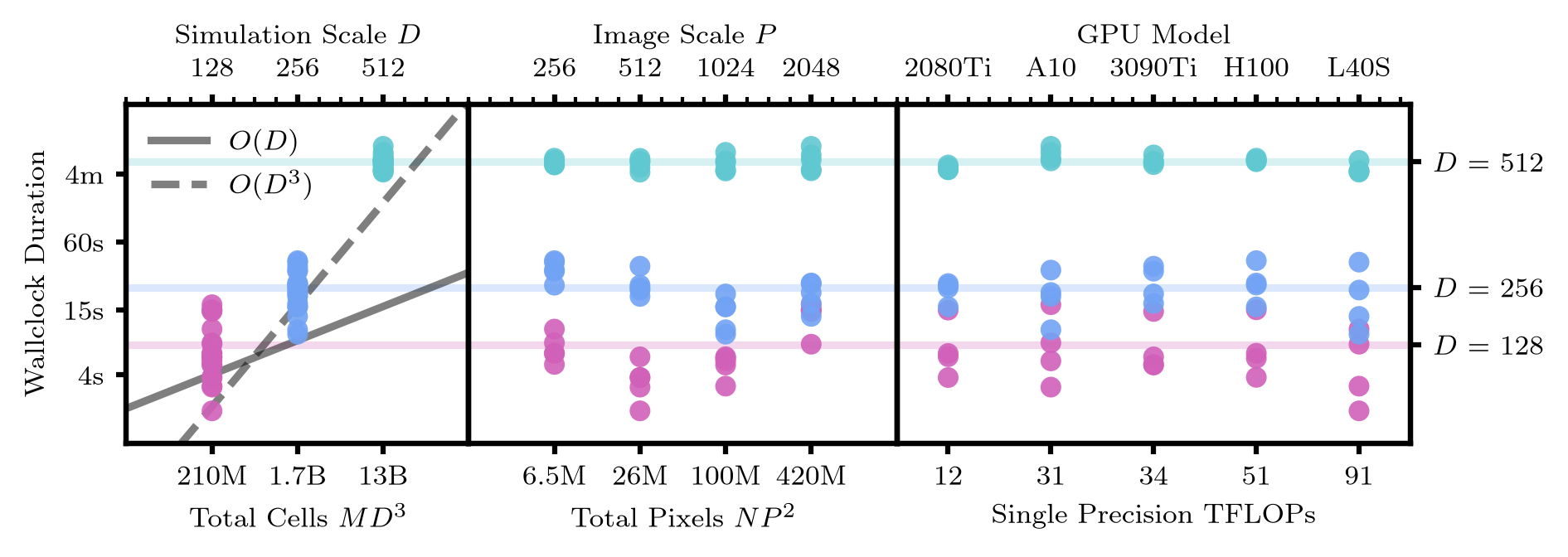}
    \caption{Durations for the backend render routine when running with lookback generating $N=100$ images (consisting of $P^2$ pixels) each sampling $M=100$ simulation snapshots (consisting of $D^3$ cells). Performance shows little dependence on both the number of pixels in the image and the GPU model used, despite a wide range of theoretical performance ceilings considered (as measured in TFLOPS). The size of simulation dataset $D$ has the most significant effect on the runtime, with duration scaling as $O(D)$ for small datasets, and $O(D^3)$ for large datasets (as the system bottlenecked by file read speed, see Figure~\ref{fig:2080ti}).}
    \label{fig:wallclocks}
\end{figure*}

\begin{figure*}
    \centering
    \includegraphics[width=2\columnwidth]{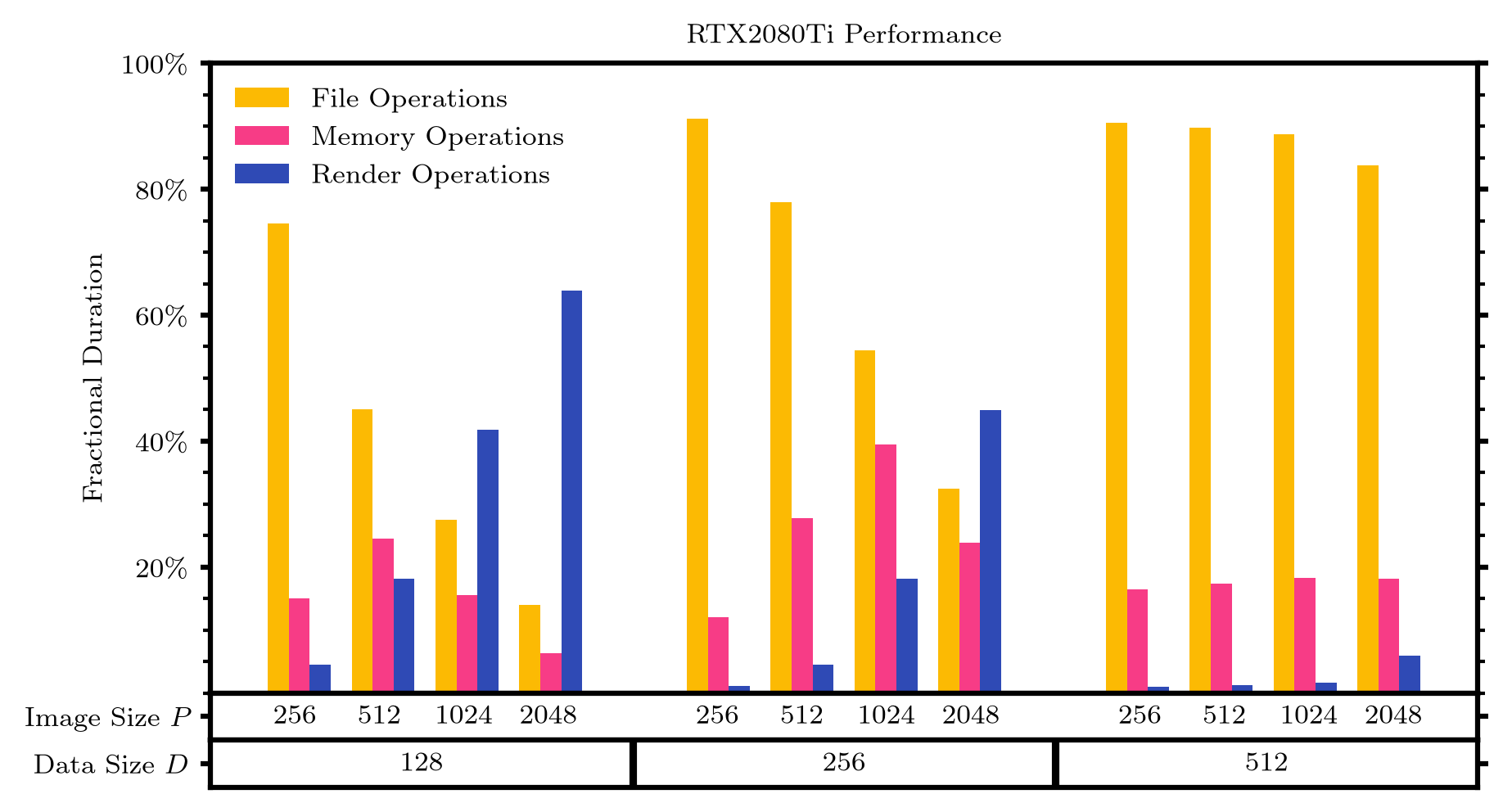}
    \caption{Fractional contribution to runtime for a series of renders performed using a single RTX2080Ti, changing the simulation size $D$ and image size $P$. All renders were performed using the lookback routine, yielding $N=100$ images each scanning $M=100$ snapshots. Small simulations ($D=128$) see comparative contributions from file, memory and render operations depending on the image size $P$, but for larger simulations, the cost of file reads completely dominates the runtime and changing the image size $P$ has negligible effect. Fractional durations are measured with respect to the wallclock duration, asynchronous device-host execution means the sum of operation fractions is not necessarily $100\%$.}
    \label{fig:2080ti}
\end{figure*}

Having discussed the possibility for local variations, we report some performance metrics for renders performed on the Institute of Science and Technology Austria's Scientific Computing Cluster. Figure~\ref{fig:wallclocks} shows the wall-clock durations for a series of tests producing a total of $N=100$ images, each sampling $M=100$ simulation snapshots. We consider the effect of changing the size of the simulation dataset $D$, image size $P$ and GPU model. 
\\~\\
Consistent with the scaling discussion of Section~\ref{sec:render_optimisation}, calculating pixel values concurrently results in a runtime that has little dependence on the size of images rendered. The main determinant for runtime ends up being the size of the domain rendered, which exhibits different scaling for small and large simulations. To better explain this trend, Figure~\ref{fig:2080ti} reports tests using a single consumer grade GPU (a RTX2080Ti), showing the fractional contribution to the total runtime for each simulation/image size. As the number of pixels in each image $P^2$ increases, the runtime fraction occupied by the render operation increases. However, this effect is only relevant for smaller simulations, for $D=512$ the runtime is entirely dominated by the cost of reading files from storage, with only minor contributions from memory copy operations and no significant dependence on the image size $P$. When the simulation size is small, and the render time occupies a significant fraction of the runtime, the total duration scales weakly with size, as the path summation algorithm has intrinsic $O(D)$ scaling (see Section~\ref{sec:render_optimisation}). However for large simulations, where the cost of reading the dataset from memory occupies the majority of the runtime, the runtime scales as $O(D^3)$ as the cost of file reads scales directly with the file's size. See Section~\ref{sec:memory_optimisation} for potential mitigation strategies for this file read bottleneck.

\section{Limitations and Future Development}
\label{sec:caveats}
\texttt{cuDART} relaxes many assumptions commonly imposed on synthetic observations of optically thin relativistic systems. Here we present the limitations relevant for the v1.0 release and discuss avenues for future development.

\subsection{Aliasing}
\label{sec:aliasing}
\begin{figure}
    \centering
    \includegraphics[width=\linewidth]{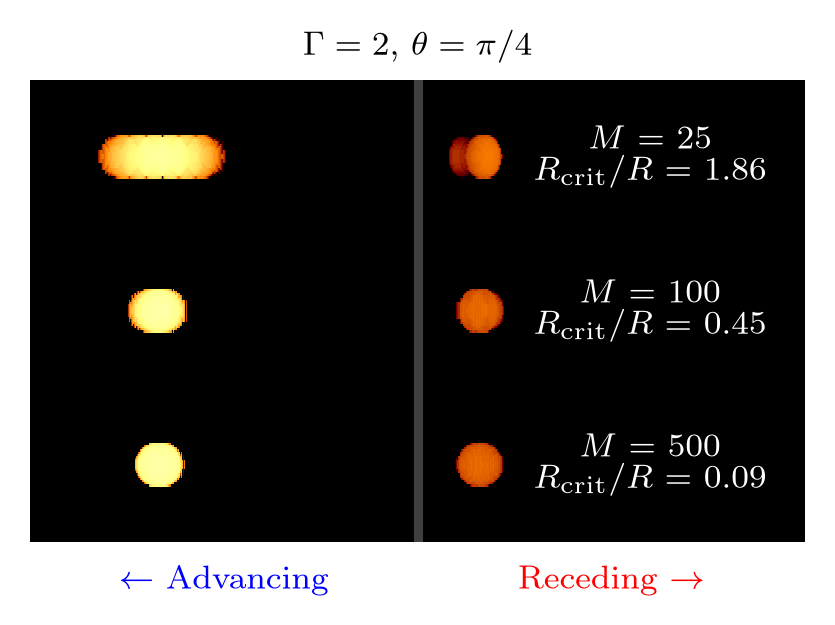}
    \caption{Comparison between renders using simulation data sampled at increasing cadences. In the top panels (low cadence), the resolved length scale $R_\mathrm{crit}$ (see Equation~\ref{eq:r_crit}) exceeds the emitter radius $R$, resulting in multiple images of the emitter. Increasing the cadence reduces $R_\mathrm{crit}$, resulting in reduced smearing and proper recovery of the observed spherical structure (to a precision set by $R_\mathrm{crit}/R$).}
    \label{fig:alias}
\end{figure}

High quality synthetic observations require simulation data that is finely sampled in both space and time. Without lookback, a single simulation snapshot is sufficient but with lookback included, multiple simulations snapshots must be read to determine the system state cross various epochs. If these snapshots are too sparsely spaced in time, the renderer will not have enough information to accurately recover the simulation state between snapshots, resulting in artificial smearing and even aliasing. This effect becomes significant when the sampling interval for the simulation state is greater than the \textit{observed} self-crossing time of an emitting element. If the user wishes to resolve a region with characteristic length scale $R$ and velocity $v$, the sampling interval $\Delta t$ should satisfy
\begin{equation}
    \label{eq:dt_alias}
    \Delta t < \Delta t_\mathrm{crit} \equiv\frac{1-\beta \cos(\theta)}{\sin (\theta)}\frac{R}{v}.
\end{equation}
Alternatively, given a sampling cadence $\Delta t$, the user can determine the minimum length scale resolvable by the rendering routine as 
\begin{equation}
    \label{eq:r_crit}
    R_\mathrm{crit} = \frac{v\sin(\theta)\Delta t}{1-\beta \cos(\theta)} = v_\mathrm{T}\Delta t.
\end{equation}
The smallest resolvable scale is then the observed distance traversed by the emitter in the interval between snapshots. Figure~\ref{fig:alias} compares renders using simulation data written with increasing cadence, for anti-parallel ejecta with radius $R$. When rendering using only a small number of snapshots $M$, the interval between snapshots is large and the resolved radius exceeds the size of the emitter, resulting in strong smearing and the appearance of multiple images (aliasing). At higher cadences, these aliased images are lost and the degree of smearing reduces. For $M=500$, the resolved radius is much smaller than the size of the emitter, resulting in accurate recovery of the observed spherical structure. Depending on the target length/velocity scale and observer orientation, the critical sampling interval may become prohibitively small, requiring a large number of simulation snapshots and presenting a significant storage burden. One method to alleviate such a burden is to render higher cadence snapshots that are traced to form partial images and then deleted, keeping only lower cadence writes for other coarse-timestep analysis. 

\subsection{Simulation Sizes}

Currently, \texttt{cuDART} is only capable of rendering data that can fit entirely within device memory, and so is limited by the size of the GPU VRAM. This does not mean that the code cannot be used to image simulation datasets that are larger than VRAM, as the user can perform partial renders using sub-regions of the full domain and sum the results. There is currently no automated support for this process, though we expect to implement this additional flexibility in future versions.

\subsection{Attenuation and Scattering}

Part of the efficiency of \texttt{cuDART} comes from its embarrassingly parallelised path summation routine, which requires no cross-talk between threads and only summation between different simulation snapshots. This prevents the system from being used to treat systems that are not optically thin, as there is no inbuilt treatment for effects such as attenuation or scattering. While in principle these effects can be included, as the primary use case for \texttt{cuDART} is optically thin emission from astrophysical radio sources, these features have been left absent from the v1.0 release. 

\subsection{General Relativisitc Effects}

While the code self-consistently accounts for Lorentzian boosting from fluid rest frames to the observer, throughout the computation all rays are assumed to be cast on straight lines through the domain. There is no support for processing non-Minkowski metrics, so the code should only be used to visualise sources that can be reasonably approximated to exist within flat spacetimes. 

\subsection{Rest-frame Emissivity and Polarisation}
\label{sec:rest-frame}
The code assumes that the emission as defined in the fluid rest frame is isotropic. In reality, that may not be the case, especially if the presence of magnetic fields impart anisotropy within the rest frame. Further, all rest-frame emission is modelled as power-law in frequency. This assumption could be relaxed provided an analytical mapping between the rest frame emissivity at a reference frequency to a more general frequency range can formed, but this is not currently implemented. Just as important, without consideration of magnetic fields the code has no way to compute the polarisation state of radiation emergent from the system. If reasonable models for rest frame polarisation can be informed, then summation of polarisation vectors along the ray line of sight could be implemented to allow for polarisation states to be included in the output. However if significant Faraday rotation is expected to take place during the integration, then as for optically thick systems, increased crosstalk is required and a new integration framework would need to be introduced. For these reasons, no treatment of magnetism or polarisation is included in the v1.0 code release. 

\section{Summary and Conclusions}
\label{sec:conclusions}
\texttt{cuDART} (CUDA/3DDDA accelerated ray tracing) is a relativistic post-processing code designed for producing synthetic observations of optically thin emission from large simulation datasets. The code automatically accounts for relativistic effects such as boosting, but also geometric effects associated with the finite time delay across astrophysical distances. The code is fully capable of visualising complex systems with inhomogeneous and dynamic emissivity and velocity fields discretised in the lab frame, requiring only
\begin{itemize}
    \item The monochromatic rest-frame emissivity
    \item The fluid velocity in units of $c$
    \item The power-law index for the rest frame emissivity
\end{itemize}
With these inputs, the code is able to produce synthetic observations from arbitrary viewpoints, observer times and frequency bands. Simulation data may be processed as a single homogenous resolution mesh, or multiple globally heterogenous but locally homogenous sub-domains, allowing for native support of simulations codes using mesh-refinement. By relaxing the need for a static integration pattern frames (commonly either the fluid or lab frame), \texttt{cuDART} requires no assumptions to be made about the discretised/continuous nature of the source. In order to accurately resolve sources with high apparent transverse motion, a large number of simulation snapshots must be processed for a single image. Ray-tracing multiple images from a series of simulation snapshots requires efficient methods of path integration: in \texttt{cuDART} this is achieved through both algorithmic and hardware acceleration, using the 3D digital differential analyser and GPU parallelisation to rapidly traverse regular Cartesian meshes. The render operation has been optimised to the extent that for most datasets, the process is entirely bottlenecked by the unavoidable cost of reading files from storage into memory.
\\~\\
This paper provides an overview of the principle functionality of \texttt{cuDART} and explains how the advanced features included within its execution are requirements for accurately rendering synthetic observations. Further information about the code usage is available via \href{https://cudart.readthedocs.io/en/latest/index.html}{ReadTheDocs}. Development for \texttt{cuDART} is use-case driven, if there are specific use cases that the reader feels would benefit from extensions to the code framework, we recommend the reader gets in contact with the author/developer by email or via \href{https://github.com/hwhitehead/cuDART}{GitHub}.

\section*{Acknowledgements}

We would like to thank Fraser Cowie, Katie Savard, James Matthews, Alex Cooper and many others for insightful discussions during development. We would especially like to thank Katie Savard for beta testing early versions of this codebase. HW acknowledges support by the Science and Technology Facilities Council Grant Number
ST/W000903/1 and the ISTA-Fellow programme at the Institute of Science and Technology Austria. ELE acknowledges funding from a Royal Society Studentship (URF\textbackslash R1\textbackslash221062). CE acknowledges a Science and Technologies Facilities Council studentship ST/X508664/1.

\section*{Data Availability}

\texttt{cuDART} is freely available on \href{https://github.com/hwhitehead/cuDART}{Github}, with documentation via \href{https://cudart.readthedocs.io/en/latest/index.html}{ReadTheDocs}. The data underlying this article will be shared on reasonable request
to the corresponding author.



\bibliographystyle{rasti}
\bibliography{citations} 




\appendix


\bsp	
\label{lastpage}
\end{document}